\documentclass[aps,twocolumn,showpacs]{revtex4}
\usepackage{graphicx}
\def\pt{p_T}
\def\v{$v_2^h(\pt,b)$}
\def\dis{distribution}

\def\bq{\begin{eqnarray}}
\def\eq{\end{eqnarray}}
\begin{document} 
\title {Inclusive Ridge Distributions in Heavy-Ion Collisions}
\author
 {Rudolph C. Hwa$^1$ and Lilin Zhu$^2$}
\affiliation
{$^1$Institute of Theoretical Science and Department of
Physics\\ University of Oregon, Eugene, OR 97403-5203, USA\\
\bigskip
$^2$Department of  Physics, Sichuan
University, Chengdu  610064, P.\ R.\ China}
\date{\today}
\begin{abstract} 
The formation of ridges induced by semihard scattering in nuclear collisions is included in the description of single-particle distributions for both pion and proton at low transverse momenta. The ridge component is characterized by an azimuthal dependent factor that is derived in the study of the ridge structure in two-particle correlation distributions involving triggers. It is shown that the inclusive ridge can reproduce the observed data on $v_2(p_T)$ if the base component underlying the ridge has no azimuthal dependence. A common description of pion and proton spectra is given in the recombination model that can smoothly join the low- and intermediate-$p_T$ regions. All the important properties of single-particle distributions in those regions can be satisfactorily described in this approach.

\pacs{25.75.-q, 25.75.Dw}
\end{abstract}
\maketitle

\section{Introduction}
As the data on single-particle distributions of identified hadrons produced in heavy-ion collisions become more abundant and precise \cite{sa, ja, ba, sa1, igb, iga, ba1}, more demands are put on theoretical models to reproduce them.  It is generally recognized that in Au-Au collisions at $\sqrt{s_{NN}}= 200$ GeV at the Relativistic Heavy-Ion Collider (RHIC) the low transverse-momentum $(p_T)$ region $(p_T < 2$ GeV/c) is well described by hydrodynamics \cite{dat} and the high-$p_T$ region ($p_T > 6$ GeV/c) by perturbative QCD \cite {mv}, both subjects being reviewed recently in Ref.\ \cite{fn}.  In the intermediate region ($2 < p_T < 6$ GeV/c) neither approaches work very well.  What stands out in that region are the large baryon/meson ratio and quark-number scaling (QNS), which give empirical support to the recombination/coalescence models \cite{hy, gkl, fmnb, hwa, hfr, mv1}.  The connection between the intermediate- and high-$p_T$ regions is smooth, since the dominance of shower-shower recombination is equivalent to parton fragmentation.  The transition across the lower $p_T$ boundary at $p_T \sim 2$ GeV/c is not so smooth because of the difference in the continuum description in hydrodynamics and the parton description in hadronization.  Our aim in this article is to extend our previous considerations \cite{chy, hz} to the lower-$p_T$ region and to describe in a self-consistent way both the $p_T$ and azimuthal $\phi$ behaviors of pions and protons without explicit reliance on hydrodynamics.

One specific point that motivates our study is related to the question of what happens to the initial system within 1 fm/c after collision.  Semihard partons created within 1 fm from the surface will have already left the initial overlap region before thermalization is complete.  There are many of them with parton transverse-momentum $k_T \sim $2-3 GeV/c even at RHIC, let alone at the Large Hadron Collider (LHC).  They are minijets that can cause azimuthal anisotropy, not accounted for by conventional hydrodynamics.  It is known that in events triggered by jets there is a ridge phenomenon in the structure of associated particles with narrow $\Delta\phi$ (azimuthal angle relative to that of the trigger) and extended $\Delta\eta$ (pseudorapidity relative to the trigger).  Such a structure should be present in the inclusive distribution even if triggers are not used to select the jet events.  When $k_T$ is low enough so that minijets are copiously produced, the corresponding effect on the $\phi$ anisotropy can become dominant, rendering the consideration of pressure gradients along different $\phi$ direction unreliable if semihard scatterings are ignored.  In this paper we give specific attention to the ridge contribution to the single-particle distributions in the low-$p_T$ region. It is in this sense that we use the terminology: inclusive ridge \dis.

Another area of concern is the variation of the $p_T$ dependence as the focus is moved to the low-$p_T$ region, where pion and proton appear empirically to have different behaviors.  In the parton recombination model the hadrons should have the same inverse slope as that of the coalescing quarks if the hadrons are formed by recombination of the thermal partons, but because of the difference in the meson and baryon wave functions, the net $p_T$ distributions turn out to be different.  This line of analysis takes into account the quark degree of freedom just before hadronization, which is overlooked by the fluid description of the flow effect.  The burden is to show that the data on $v_2(p_T)$ can be reproduced for both pion and proton at low $p_T$ without the hydro description of elliptic flow.  That is indeed what we shall show for various centralities.  

We confine our consideration in this paper to the physics at midrapidity.  At larger $\eta$ there are other issues, such as large $p/\pi$ ratio \cite{iga} and large $\Delta\eta$ distribution of triggered ridge \cite{ba1}, which have been examined in Refs. \cite{hz1, ch}, and will not be further considered here.

\section{Single-particle distribution with ridge }
We begin with a recapitulation of our description of single-particle distribution \cite{hy, chy, hz}.  At low $p_T$ we consider only the recombination of thermal partons, so the pion and proton spectra at $y=0$ are given by
\begin{eqnarray}
p^0 {dN^{\pi}\over dp_T} &=& \int \prod_{i=1}^2 \left[{dq_i\over q_i} {\cal T} (q_i)\right] {\cal R}^{\pi}(q_1,q_2,p_T),     \label{1} \\
p^0 {dN^p\over dp_T} &=& \int \prod_{i=1}^3 \left[{dq_i\over q_i} {\cal T} (q_i)\right] {\cal R}^p(q_1,q_2,q_3,p_T),     \label{2}
\end{eqnarray}
where ${\cal T} (q_i)$ is the thermal distribution of the quark (or antiquark) with momentum $q_i$, and ${\cal R}^h$ is the recombination function (RF) for $h=\pi$ or $p$.  On the assumption that collinear quarks make the dominant contribution to the coalescence process (so that the integrals are one-dimensional for each quark along the direction of the hadron), the RFs are
\begin{eqnarray}
{\cal R}^{\pi}(q_1,q_2,p) &=& {q_1q_2\over p^2} \delta\left(\sum_{i=1}^2 {q_i\over p} - 1 \right),    \label{3} \\
{\cal R}^p(q_1,q_2,q_3,p) &=& f\left({q_1\over p},{q_2\over p},{q_3\over p} \right) \delta\left(\sum_{i=1}^3 {q_i\over p} - 1 \right)     \label{4}
\end{eqnarray}
where the details of $f(q_i/p)$ that depends on the proton wave function are given in \cite{hy}, and need not be repeated here.  The main point to be made here is that if the thermal distribution ${\cal T}(q_i)$ has the canonical invariant form
\begin{eqnarray}
{\cal T}(q) = q{dN^q\over dq} = Cqe^{-q/T},     \label{5}
\end{eqnarray}
then the $\delta$-functions in the RFs require that $dN^h/p_Tdp_T$ has the common exponential factor, $\exp(-p_T/T)$, for both $h=\pi$ and $p$.  The prefactors are different; we simply write down the results obtained previously
\begin{eqnarray}
{dN^{\pi}\over p_Tdp_T} &=& {\cal N}_{\pi}e^{-p_T/T},     \label{6}  \\
{dN^p\over p_Tdp_T} &=& {\cal N}_p{p_T^2\over m_T}e^{-p_T/T}, \quad  m_T = (p_T^2 + m_p^2)^{1/2},     \label{7}
\end{eqnarray}
where ${\cal N}_{\pi} \propto C^2$ and ${\cal N}_p \propto C^3$, and $C$ has the dimension (GeV)$^{-1}$.  Note that the factor $p_T^2/m_T$ in the proton spectrum (that must be present for dimensional reason) causes the $p/\pi$ ratio to vanish as $p_T \rightarrow 0$ on the one hand, but to become large, as $p_T$ increases, on the other.  When $p_T$ exceeds 2 GeV/c, shower partons become important and the above description must be supplemented by thermal-shower ({\bf TS}) recombination that limits the increase of the $p/\pi$ ratio to a maximum of about 1 \cite{hy}.

We restrict our consideration to $p_T < 2$ GeV/c, but now broaden it to include $\phi$ dependence.  For non-central collisions the almond-shaped initial configuration leads to $\phi$ anisotropy.  The conventional description in terms of hydrodynamics relates the momentum anisotropy to the variation of pressure gradient at early times upon equilibration \cite{kh}.  The success in obtaining the large $v_2$ as observed gives credibility to the approach.  We adopt an alternative approach and justify our point of view on the basis that we can also reproduce the empirical $v_2$, as we shall show.  Furthermore, aside from offering a smooth connection with the intermediate $p_T$ region by the inclusion of {\bf TS} recombination, our approach describes also the effect of semihard scattering on the soft sector.  The ridge phenomenon that we attribute to that effect can be with trigger \cite{ch, ch1, ch2} or without trigger \cite{rch, hwa, chy, hz}.  Although data on the ridge structure must necessarily make use of triggers  in order to distinguish it from background \cite{ja1, bia, ha, ba2}, inclusive distribution must include ridges along with background.  Thus theoretically a single-particle distribution should have a ridge component in the soft sector due to undetected semihard or hard partons.  That component has $\phi$ dependence that can be calculated from geometrical consideration \cite{hz}, and has been shown to be consistent with the dependence of the ridge yield in two-particle correlation on the trigger angle $\phi_s$ relative to the reaction plane \cite{ha}.

Let us use $\rho_1^h(p_T,\phi,b)$ to denote the single-particle distribution of hadron $h$ produced at mid-rapidity in heavy-ion collision at impact parameter $b$,  i.e.
\begin{eqnarray}
\rho_1^h(p_T,\phi,b) = {dN^h\over p_Tdp_Td\phi}(N_{\rm part}),     \label{8}
\end{eqnarray}
where $N_{\rm part}$ is the number of participants related to $b$ in a known way through Glauber description of nuclear collision \cite{mrss}.  At low $p_T$ let $\rho_1^h$ be separated into two components
\begin{eqnarray}
\rho_1^h(p_T,\phi,b) = B^h(p_T,b) + R^h(p_T,\phi,b),     \label{9}
\end{eqnarray}
where $B^h(p_T,b)$ is referred to as Base, not to be confused with the bulk that is usually determined in hydrodynamics; this is a change from earlier nomenclature \cite{hz}, where the use of ``bulk" did lead to some misunderstanding.  Our emphasis here is that $B(p_T,b)$ is independent of $\phi$.  In our approach we regard the semihard partons created near the surface, and directed outward, give rise to all the $\phi$ dependence of the medium before equilibrium is established; the recoil partons being directed inward are absorbed and randomized.  The component expressed by $R^h(p_T,\phi,b)$ is referred to as ridge on the basis of its $\phi$ dependence discussed below.  The $B^h(p_T,b)$ component consists of all the soft and semihard partons that are farther away from the surface and are unable to lead to hadrons with distinctive $\phi$ dependence.  Thus the separation between $B^h(p_T,b)$ and $R^h(p_T,\phi,b)$ relies primarily on the $\phi$ dependence that the ridge component possesses.

In Ref.\ \cite{hz} we have given an extended derivation of what that $\phi$ dependence is.  It is embodied in $S(\phi,b)$ that is the segment of the surface through which a semihard parton can be emitted to contribute to a ridge particle at $\phi$.  From the geometry of the initial ellipse (with width $w$ and height $h$ that depend on $b$) and from the angular constraint between the semihard parton and ridge particle prescribed by a Gaussian width $\sigma$ determined earlier in treating the ridge formation for nuclear density not too low \cite{ch2}, it is found that
\begin{eqnarray}
S(\phi,b) = h[E(\theta_2,\alpha) - E(\theta_1,\alpha)],     \label{10}
\end{eqnarray}
where $E(\theta_i,\alpha)$ is the elliptic integral of the second kind with $\alpha=1-w^2/h^2$ and
\begin{eqnarray}
\theta_i = \tan^{-1} \left({h\over w}\tan\phi_i \right), \quad \phi_1 = \phi - \sigma, \quad \phi_2 = \phi + \sigma,     \label{11}
\end{eqnarray}
for $\phi_i \leq \pi/2$, and an analytic continuation of it for $\phi_2 > \pi/2$.  Thus $S(\phi,b)$ is completely calculable for any given $b$, and $R^h(p_T,\phi,b)$ is proportional to it.

We can now rewrite Eq.\ (\ref{9}) unambiguously as
\begin{eqnarray}
\rho_1^h(p_T,\phi,b) = B^h(p_T,b) + {S(\phi,b)\over \bar S (b)} \bar R^h(p_T,b)     \label{12}
\end{eqnarray}
where
\begin{eqnarray}
\bar S (b) = (2/\pi) \int_0^{\pi/2} d\phi S(\phi,b)    \label{13}
\end{eqnarray}
and $\bar R^h(p_T,b)$ is a similar average of $R^h(p_T,\phi,b)$.  According to Eqs.\ (\ref{6}) and (\ref{7}) the inclusive distributions $\bar \rho_1^h(p_T,b)$ should share the common exponential factor $\exp(-p_T/T)$, for $h = \pi$ and $p$, as for quarks.  That does not take into consideration the enhancement of pions at very small $p_T$ due to resonance decay.  We account for it by a phenomenological term $u(p_T,b)$, and write
{\begin{eqnarray}
\bar \rho_1^{\pi}(p_T,b) &=& {\cal N}_{\pi} (b)[1 + u(p_T,b)]e^{-p_T/T},     \label{14} \\
\bar \rho_1^p(p_T,b) &=& {\cal N}_p(b){p_T^2\over m_T}e^{-p_T/T},     \label{15}
\end{eqnarray}
where the resonance effect on the proton is neglected because of baryon-number conservation.  These expressions are for the left-hand side of Eq.\ (\ref{12}) after $\phi$ averaging.  The base term $B^h(p_T,b)$ on the right side is the soft component without the contribution from semihard scattering near the surface and should have the same common structure as in Eqs.\ (\ref{6}) and (\ref{7}) due to thermal parton recombination, except that the inverse slope is lower without the enhancement by the energy loss from the semihard partons.  We can therefore write
\begin{eqnarray}
B^{\pi}(p_T,b) &=& {\cal N}_{\pi}(b)[1 + u(p_T,b)]e^{-p_T/T_B},     \label{16} \\
B^p(p_T,b) &=& {\cal N}_p(b){p_T^2\over m_T}e^{-p_T/T_B},    \label{17}
\end{eqnarray}
where $T_B < T$.  It then follows that
\begin{eqnarray}
\bar R^{\pi}(p_T,b) &=& {\cal N}_{\pi}(b)[1+u(p_T,b)] \bar R_0(\pt)     \label{18} \\
\bar R^p(p_T,b) &=& {\cal N}_p(b){p_T^2\over m_T} \bar R_0(\pt),     \label{19}
\end{eqnarray}
where
\begin{eqnarray}
\bar R_0(\pt)=e^{-\pt/T}-e^{-\pt/T_B}=e^{-p_T/T_B}(e^{p_T/\tilde{T}}-1) \quad    \label{19a}  \\
{1\over \tilde{T}} = {1\over T_B} - {1\over T} = {\Delta T\over T_BT}, \qquad  \Delta T = T - T_B.  \qquad   \label{20}
\end{eqnarray}
There are two undetermined inverse-slopes: $T_B$ and $T$, common for both $\pi$ and $p$.  They are for single-particle inclusive distributions, so only $T$ is directly observable.  We postpone phenomenology to a later section.  In ridge analysis using triggered events for two-particle correlation the two corresponding inverse slopes are separately measured \cite{bia}.  Here, however, we are dealing with single-particle distributions.  The difference between $T_B$ and $T$ has to do with ridges and their effect on the $\phi$ distribution.  Thus we expect $\Delta T$ to be related to azimuthal asymmetry, a topic we next turn to.

\section{Quadrupole Moments of $\phi$ Asymmetry}
This topic is usually referred to as elliptic flow, a terminology that is rooted in hydrodynamics.  Since we have not used hydro in the previous section, it is more appropriate to use the unbiased language initiated in Ref.\ \cite{tk}, and call it azimuthal quadrupole.  It is the familiar $v_2$ that is defined by
\begin{eqnarray}
v_2^h(p_T,b) = \langle \cos2\phi \rangle_{\rho_1}^h = {\int_0^{2\pi} d\phi \cos2\phi\rho_1^h(p_T,\phi,b)\over \int_0^{2\pi} d\phi\rho_1^h(p_T,\phi,b)}.     \label{21}
\end{eqnarray}
Using Eqs.\ (\ref{12}) - (\ref{20}) yields
\begin{eqnarray}
v_2^h(p_T,b) &=& {[2\bar R(\pt,b)/\pi\bar S(\pt,b)]\int_0^{\pi/2} d\phi \cos2\phi S(\phi,b) \over B(\pt,b)+\bar R(\pt,b)}   \nonumber \\
&=& {\langle \cos2\phi \rangle_S\over Z^{-1}(p_T) + 1},     \label{22}
\end{eqnarray}
where
\begin{eqnarray}
\langle \cos2\phi \rangle_S &=& {2/\pi\over \bar S (b)} \int_0^{\pi/2} d\phi \cos2\phi S(\phi,b),     \label{23} \\
Z(p_T) &=& e^{p_T/\tilde T} - 1.     \label{24}
\end{eqnarray}
These equations are remarkable in that the $b$ dependence resides entirely in Eq.\ (\ref{23}) and the $p_T$ dependence entirely in Eq.\ (\ref{24}); furthermore, there is no explicit dependence on the hadron type nor the resonance term represented by $u(\pt,b)$.  As we have noted at the end of the preceding section, $T$ can be determined by the $p_T$ spectra, but $T_B$ is not directly observable.  However, the quadrupole is measurable, so it can constrain $\tilde T$ and therefore $T_B$.  In short, the two parameters $T$ and $T_B$ can be fixed by fitting the data on $\bar \rho_1^h(p_T,b)$ and $v_2^h(p_T,b)$.  

Without using a model to describe the evolution of the dense medium, it is clear that we cannot predict the values of $T$ and $T_B$.  
However, our aim is to discover how far one can go without using such a model. 
Neither $T$ nor $T_B$ depend on $\phi$.  Yet non-trivial $v_2^h(p_T,b)$ can be obtained because of the presence of the ridge term in Eq.\ (\ref{12}).  If phenomenology turns out to support this interpretation of azimuthal asymmetry, as we shall do in the next section, then the ridges induced by undetected semihard partons play a more important role in giving rise to the $\phi$ dependence in inclusive single-particle distribution than hydro expansion that is based on assuming equilibration to be completely at a later time without semihard scattering.

From Eqs.\ (\ref{10}) and (\ref{23}) we can calculate $\left< \cos 2\phi\right>_S$ and obtain its dependence on $b$. For the initial elliptical configuration the width and height are 
\bq
w=1-b/2,  \qquad  h=(1-b^2/4)^{1/2} ,   \label{25}
\eq
where all lengths are in units of the nuclear radius $R_A$. Setting the Gaussian width $\sigma$ between the azimuthal angle $\phi_1$ of the semihard parton and $\phi_2$ of the ridge particle to be $\sigma=0.33$ \cite{ch2}, we determine $\left< \cos 2\phi\right>_S$ as shown in Fig.\ 1(a).

\begin{figure}[tbph]
\includegraphics[width=.4\textwidth]{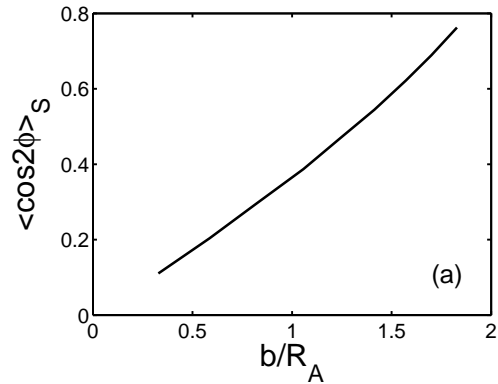}
\includegraphics[width=.5\textwidth]{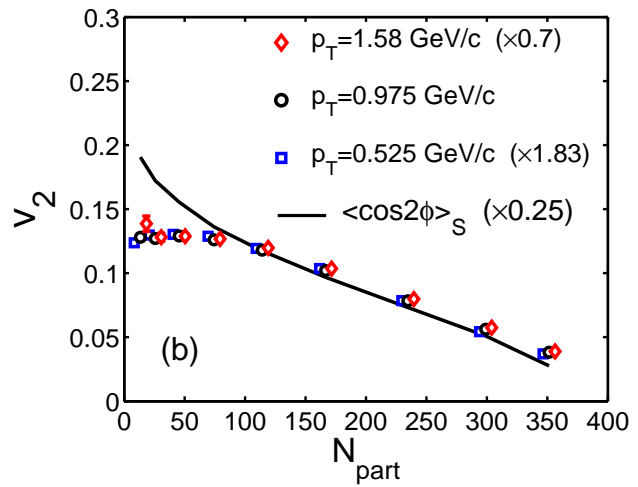}
\caption{(Color online) (a) Average of $\cos2\phi$ weighted by $S(\phi,b)$ vs  impact parameter in units of $R_A$. (b) Common dependence of $v_2^h(\pt,b)$ on $N_{part}$ for various $\pt$, shifted vertically for comparison. The diamond and square points are horizontally shifted slightly from the points in circles to aid visualization. The solid line is from $\left<\cos2\phi\right>_S$ shown in (a), but rescaled and plotted in terms of $N_{part}$. The data are from Ref.\ \cite{ja}.}
\end{figure}

According to Eq.\ (\ref{22}) $\left< \cos2\phi\right>_S$ contains all the $b$ dependence of $v_2^h(\pt,b)$ for any $\pt$ in the soft region. To check how realistic that is phenomenologically, we show first in Fig.\ 1(b) the data on $v_2^h(\pt,N_{part})$ for three $\pt$ values from Ref.\ \cite{ja}, but shifted vertically so that they agree with the data for $\pt=0.975$ GeV/c for most of large $N_{part}$. The diamond and square points are slightly shifted horizontally to spread out the overlapping points for the sake of visual distinguishability. The fact that their dependencies on $N_{part}$ are so nearly identical is remarkable in itself. The solid line is a reproduction of the curve in Fig.\ 1(a) but plotted in terms of $N_{part}$, and reduced in normalization by a factor 0.25 to facilitate the comparison with the data points. For $N_{part}>100$ the line agrees with the data on $v_2$ very well, thus proving the factorizability of $\pt$ and $b$ dependencies of Eq.\ (\ref{22}). For $N_{part}<100$, corresponding to $b/R_A>1.3$ or centrality $>40$\%, there is disagreement which is expected because the density is too low in peripheral collisions to justify the simple formula in Eq.\ (\ref{22}). A density-dependent correction is considered in Ref.\ \cite{hz}, but will not be repeated here. Our focus in this paper is on the inclusive ridge, so we proceed to phenomenology on the basis that the formalism given above is valid for central and mid-central collisions at $N_{part}>100$. To have a compact analytic expression for $S(\phi,b)$ as given in Eqs.\ (\ref{10}) and (\ref{11}) to summarize the $\phi$ dependence is not only economical, but also provides a succinct feature to distinguish the ridge from the base components in Eq.\ (\ref{12}). 

\section{Phenomenology}

We now determine the parameters in our model through phenomenology. A success in fitting all the relevant data can give support to our approach that emphasizes issues not considered in the standard model \cite{rv}.

Our first task is to determine the inverse slope $T$ that is shared by ${\cal T}(q), \bar\rho_1^\pi(\pt,b)$ and 
$\bar\rho_1^p(\pt,b)$. Since the normalization factors in Eqs.\ (\ref{5}), (\ref{14}) and (\ref{15}) have not yet been specified, we consider first a particular centrality, 20-30\%, and fit the $\pt$ dependence of the proton spectrum for $\pt<2$ GeV/c, as shown in Fig.\ 2, and obtain
\bq
T=0.283\ {\rm GeV}.  \label{26}
\eq
Note that the one-parameter fit (apart from normalization) is very good compared to the data from Ref.\ \cite{sa}. It demonstrates that the proton is produced in that $\pt$ range by thermal partons and that the flattening of the spectrum at low $\pt$ is due to the prefactor $\pt^2/m_T$ arising from the proton wave function. 

\begin{figure}[tbph]
\includegraphics[width=.45\textwidth]{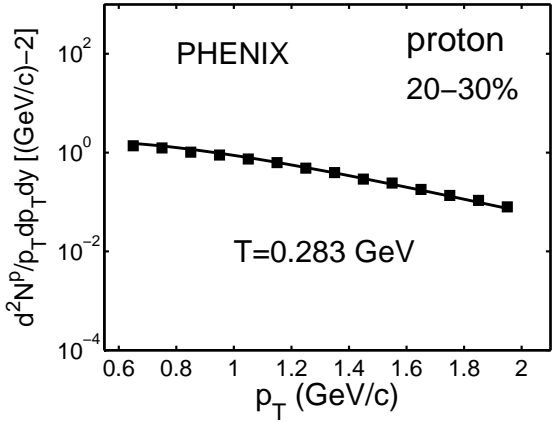}
\caption{Proton spectrum at $y\approx 0$ averaged over $\phi$ (hence, no $1/2\pi$ factor) at 20-30\% centrality. The solid line is a fit of the data by Eqs.\ (\ref{15}) and (\ref{26}) with free adjustment of normalization. The data are from Ref.\ \cite{sa}.}
\end{figure}

Having determined $T$, we next consider the pion spectrum $\bar\rho_1^\pi(\pt,b)$. According to Eq.\ (\ref{14}) it has the same exponential factor as does $\bar\rho_1^p(\pt,b)$, but has also an additional factor $[1+u(\pt,b)]$ due to resonance decay. We show in Fig.\ 3 the data from PHENIX \cite{sa} on the pion \dis\ for 20-30\% centrality; the $\exp(-\pt/T)$ factor is shown by the dashed line, the normalization being adjusted to fit (and to be discussed later). For $\pt>1$ GeV/c they agree very well, demonstrating the validity of the common $T$. For $\pt<1$ GeV/c there is resonance contribution to the pion spectrum which we cannot predict. Thus we fit the low-$\pt$ region by the addition of a term $\exp(-\pt/T_r)$, shown by the dash-dotted line, corresponding to $T_r=0.174$ GeV. The sum depicted by the solid line agrees with the data perfectly. The point of this exercise is mainly to show that the common $\exp(-\pt/T)$ behavior is valid for pion as for proton, but the reality of resonance contribution for $\pt<1$ GeV/c obscures that commonality. Converting the resonance term to the form given in Eq.\ (\ref{14}) we write
\bq
u(\pt,b)=u_0(b) e^{-\pt/T_0} ,  \label{27}
\eq
where $T_0=0.45$ GeV and $u_0=3.416$ for 20-30\% centrality. We do not regard this $u$ term as a fundamental part of our model; we attach the  factor $[1+u(\pt,b)]$ to all expressions of the pion \dis s, as in Eqs.\ (\ref{16}) and (\ref{18}). Of more significance is the role that $T$ has played in the phenomenology, and so far $T_B$ has played no role.

\begin{figure}[tbph]
\includegraphics[width=.45\textwidth]{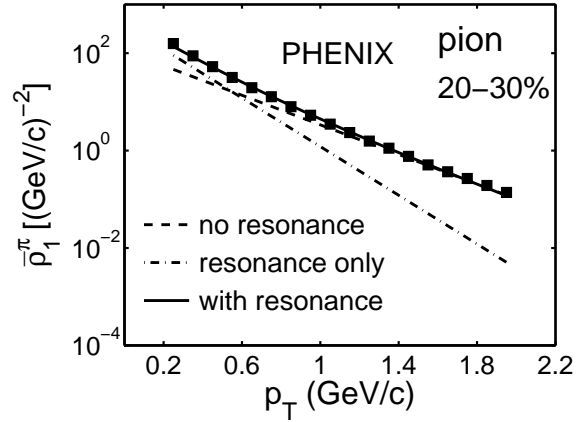}
\caption{Pion spectrum showing $e^{-\pt/T}$ by the dashed line, and the resonance contribution  by the dash-dotted line. The sum is in solid line. The data are from Ref.\ \cite{sa}.}
\end{figure}

$T_B$  is not directly related to any observable spectrum, since it describes the $\pt$ dependence of the base $B^h(\pt,b)$ that lies under the ridge. The important concept we advance here is that it is $\phi$ independent, and that $R^h(\pt,\phi,b)$ carries all the $\phi$ dependence. Thus we turn to $v_2^h(\pt,b)$ in Eq.\ (\ref{22}) and examine its $\pt$ dependence for both $h=\pi$ and $p$. In order to emphasize the universality between $\pi$ and $p$, we consider $v_2^h$ versus the transverse kinetic energy $E_T$, for $E_T<0.8$ GeV, where
\bq
E_T(p_T)=m_T(\pt)-m_h .   \label {28}
\eq
We adopt the ansatz that $\pt$ is to be replaced by $E_T$ in Eq.\ (\ref{24}) so as to account for the mass effect, i.e.,
\bq
Z(\pt)=e^{E_T(\pt)/\tilde T}-1 ,  \label{29}
\eq
where $\tilde T$ is as given in Eq.\ (\ref{20}). In Fig.\ 4 is shown the data from Ref.\ \cite{ja} when $v_2^h$ is plotted against $E_T$ for 20-30\% centrality. We fit the data points for both $h=\pi$ and $p$ by Eqs.\ (\ref{22}) and (\ref{29}) with the choice 
\bq
T_B=0.253\pm 0.003\ {\rm GeV} ,  \label{30}
\eq

\begin{figure}[tbph]
\vspace*{-.5cm}
\includegraphics[width=.45\textwidth]{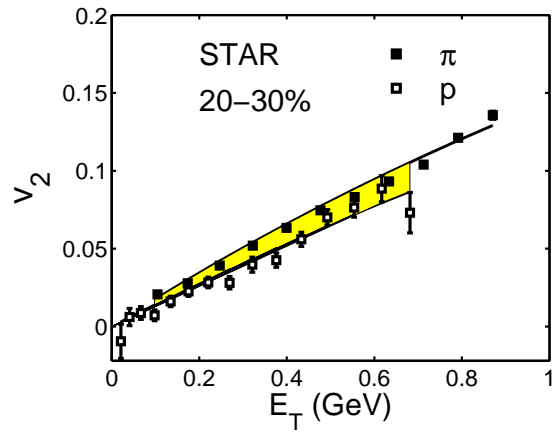}
\caption{(Color online) $v_2^h$ for $h=\pi$ and $p$. The shaded region corresponds to $T_B=0.253\pm0.003$ GeV. The data are from Ref.\ \cite{ja}.}
\end{figure}

\noindent which is represented by the shaded region in Fig.\ 4. The upper boundary of that region is for $T_B=0.25$ GeV that fits the pion $v_2$ almost perfectly, and the lower boundary is for $T_B=0.256$ GeV that fits well the proton $v_2$. 
It is evident that $v_2$ is very sensitive to $T_B$ due to the exponential factor in Eq.\ (\ref{29}), yet the data support a common value for $T_B$ to within 1-2\% deviation for pion and proton production.
One cannot expect an accuracy better than that in the universality of $v_2^h$ for $h=\pi$ and $p$. We regard this result to be remarkable, since the normalization of $v_2^h$ is fixed by Eq.\ (\ref{22}) without freedom of adjustment. Note that we have not used any more parameters besides $T$ and $T_B$ to accomplish this, which is a fitting procedure not more elaborate than the hydro approach where the initial condition and viscosity are adjusted.

So far we have concentrated on 20-30\% centrality partly because we want to separate the $\pt$ and $\phi$ dependencies from the issue of centrality dependence, and partly because $v_2^h(\pt,b)$ is large at 20-30\% centrality for low $\pt$. To extend our consideration to other centralities, we fix $T$ and $T_B$ at the values obtained in Eqs.\ (\ref{26}) and (\ref{30}) so that $Z(\pt)$ is no longer adjustable. The centrality dependence of \v\ is then examined using Eq.\ (\ref{22}). Figure 5 shows the results for different centrality bins for both $h=\pi$ and $p$. The shaded regions due to the uncertainty in Eq.\ (\ref{30}) become narrower in more central collisions. The agreement with data from STAR \cite{ja} is evidently very good. Since there has been no more adjustment of free parameters to achieve that, we find substantial support from Fig.\ 5 for our view that the $\phi$ dependence arises entirely from the ridge component in the inclusive \dis. This raises serious question on whether viscous hydrodynamics is the only acceptable description of heavy-ion collisions, if the reproduction of \v\ is the primary criterion for the success of a model.

\begin{figure}[tbph]
\hspace*{-.5cm}
\includegraphics[width=.5\textwidth]{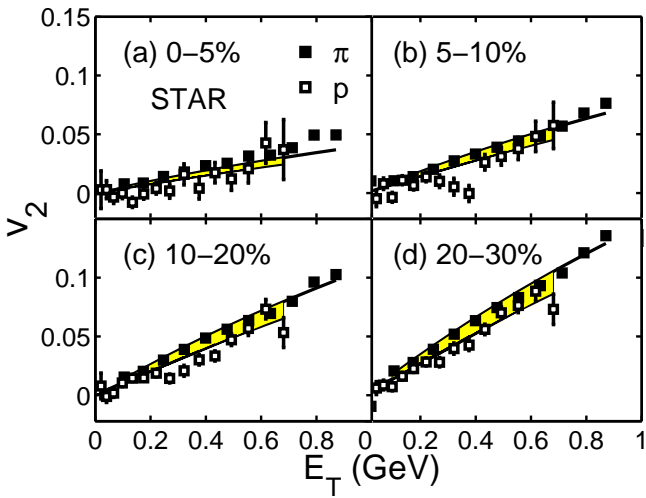}
\caption{(Color online) Same as in Fig.\ 4 for four centrality bins.}
\end{figure}

It is possible to further improve the agreement between the values of $v_2^h$ for pion and proton in Fig.\ 5 if those figures are replotted in accordance to the idea of quark number scaling (QNS), i.e., $v_2^h/n_h$ vs $E_T/n_h$, where $n_h$ is the number of constituent quarks in hadron $h$ \cite{mv1,kcg}. As we have considered QNS and its breaking in the recombination model before \cite{chy}, we do not revisit that problem here, especially since our main goal to use $v_2$ to constrain $T_B$ has already been accomplished.

Having obtained the correct centrality dependence of \v\ that is calculable, we now consider the centrality dependence of the inclusive spectra $\bar\rho_1^h(\pt,b)$. We note that the unknown
 normalization factors ${\cal N}_\pi(b)$ and ${\cal N}_p(b)$ in Eqs.\ (\ref{14}) and (\ref{15}) never enter into the calculation of \v\ because of cancellation, but for $\bar\rho_1^h(\pt,b)$ they must be reckoned with. As remarked after Eq.\ (\ref{7}), ${\cal N}_\pi(b)$ and ${\cal N}_p(b)$
are proportional to $C^2$ and $C^3$, respectively,  due to $q\bar q$ and $qqq$ recombination. The magnitude $C$ of the thermal partons depends on $b$ in a way that cannot be reliably calculated. By phenomenology on the pion spectrum it was previously estimated for $\pt>1.2$ GeV/c \cite{hz}, but that is inadequate for our purpose here; moreover, ${\cal N}_\pi(b)$ and ${\cal N}_p(b)$ have different statistical factors that can depend on $b$ because of resonances. We give here direct parametrizations of the normalization factors in terms of $N_{part}$
\bq
{\cal N}_\pi(N_{part})&=&0.516 N_{part}^{1.05} ,  \label{31}  \\
{\cal N}_p(N_{part})&=&0.149 N_{part}^{1.18} ,  \label{32}  \\
u_0(N_{part}) &=& 2.8+0.003 N_{part} . \label{33}
\eq
The parameters are determined by fitting the centrality dependence to be shown, but the essence of our prediction is in $\pt$ and $\phi$ dependencies that are not adjustable. Using the above in Eqs.\ (\ref{14}) and (\ref{15}) we obtain the curves in Fig.\ 6 (a) pion and (b) proton for three centrality bins. They agree with the data from PHENIX \cite{sa} very well over a wide range of low $\pt$. In all those curves $T$ is kept fixed at 0.283 GeV, thus reaffirming our point that both pions and protons are produced by the same set of thermal partons despite the apparent differences in the shapes of their $\pt$ dependencies.

\begin{figure}[tbph]
\includegraphics[width=.45\textwidth]{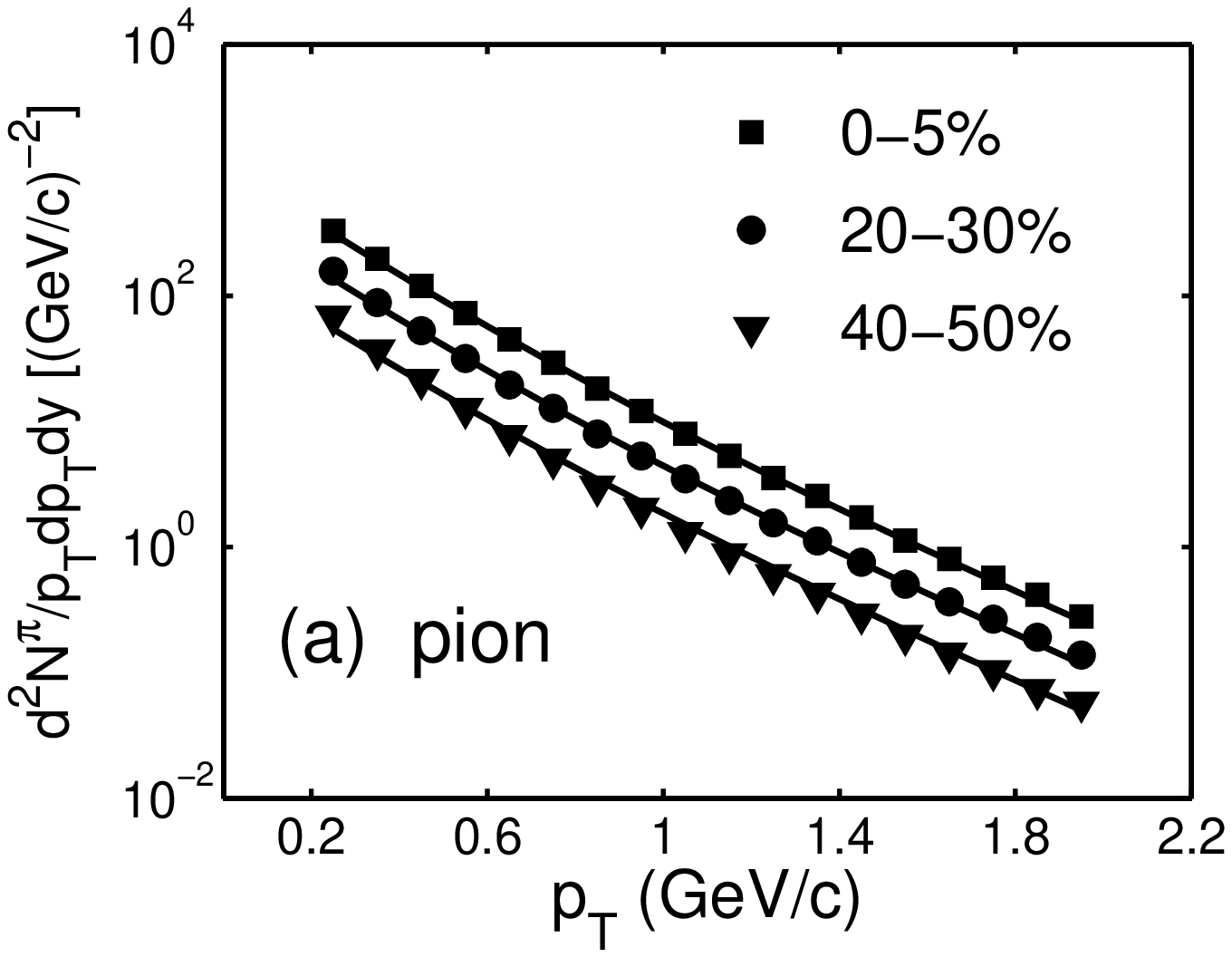}
\hspace{.5cm}
\includegraphics[width=.45\textwidth]{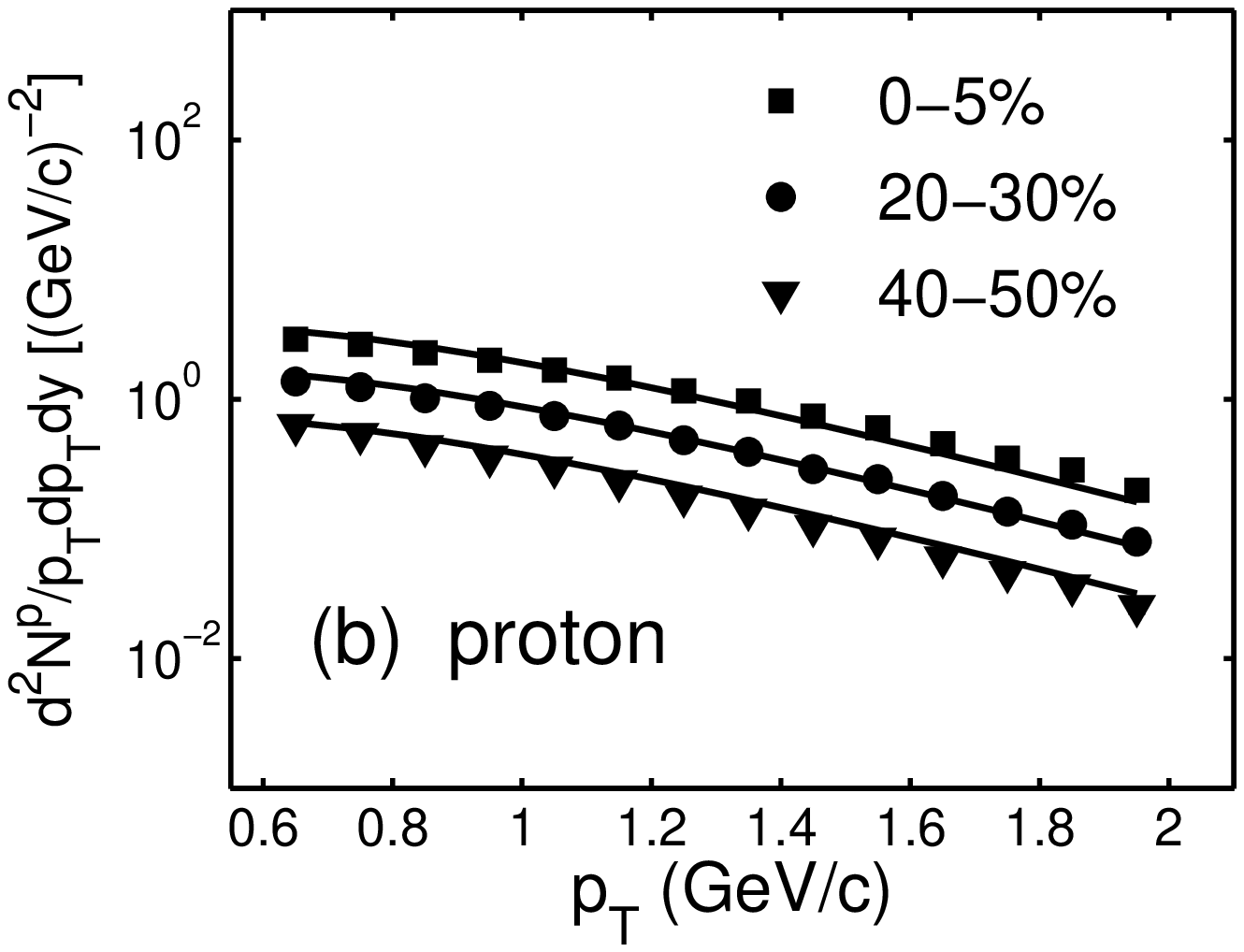}
\caption{Inclusive spectra at three centralities for (a) pion and  (b) proton. The data are from Ref.\ \cite{sa}.}
\end{figure}

\section{Inclusive Ridge Distribution}

It is now opportune for us to revisit the two-component description of the single-particle \dis\ and focus on the ridge component, in particular. As stated explicitly in Eq.\ (\ref{12}), the  $\phi$ dependence separates the $B^h(\pt,b)$ and $\bar R^h(\pt,b)$ components, the former being described by Eqs.\ (\ref{16}) and (\ref{17}), the latter by Eqs.\ (\ref{18}) and (\ref{19}). Upon averaging over $\phi$, we have
\bq
\bar\rho_1^h(\pt,b)=B^h(\pt,b)+\bar R^h(\pt,b) .  \label{34}
\eq

\begin{figure}[tbph]
\includegraphics[width=.45\textwidth]{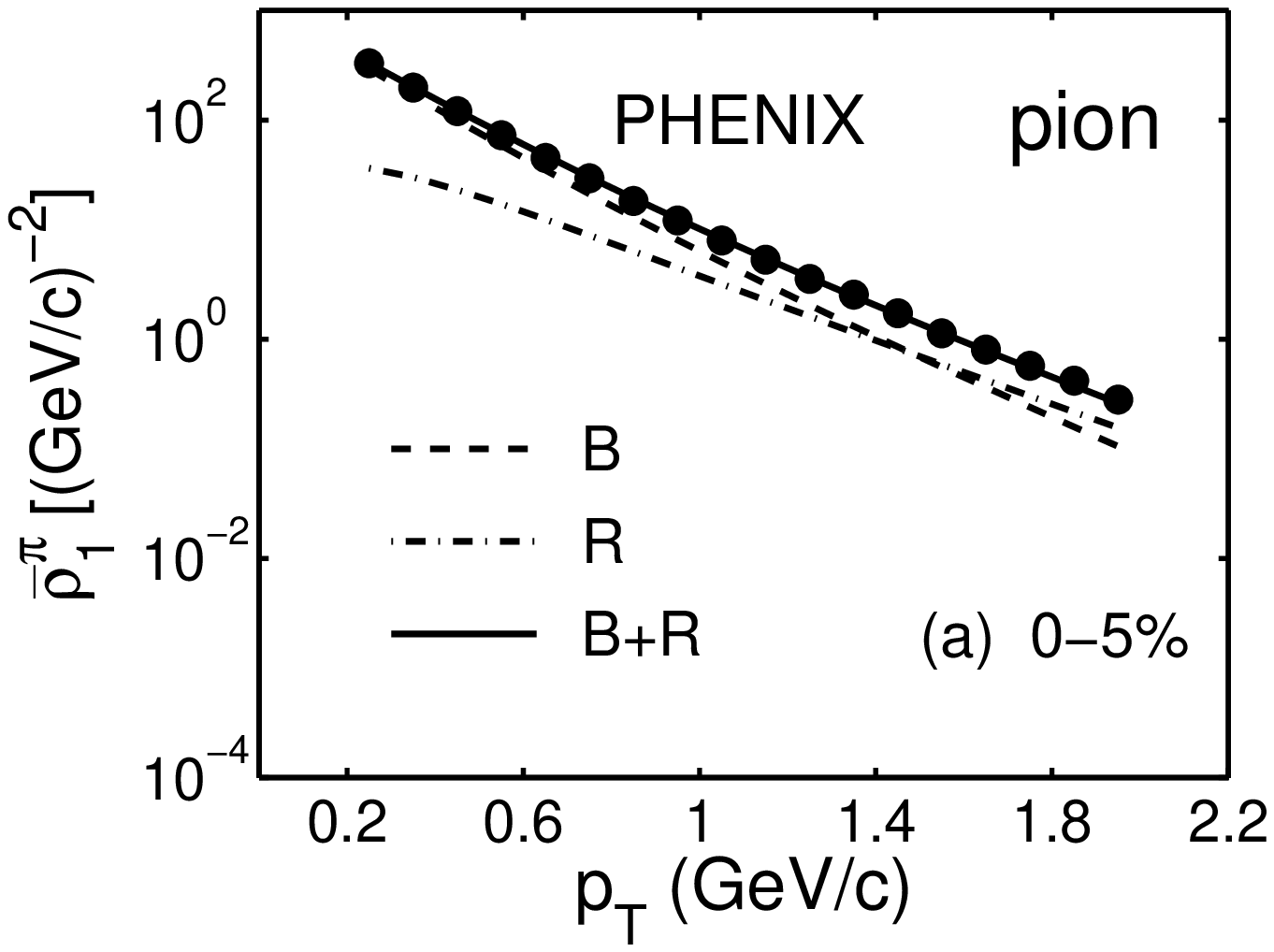}
\hspace{.5cm}
\includegraphics[width=.45\textwidth]{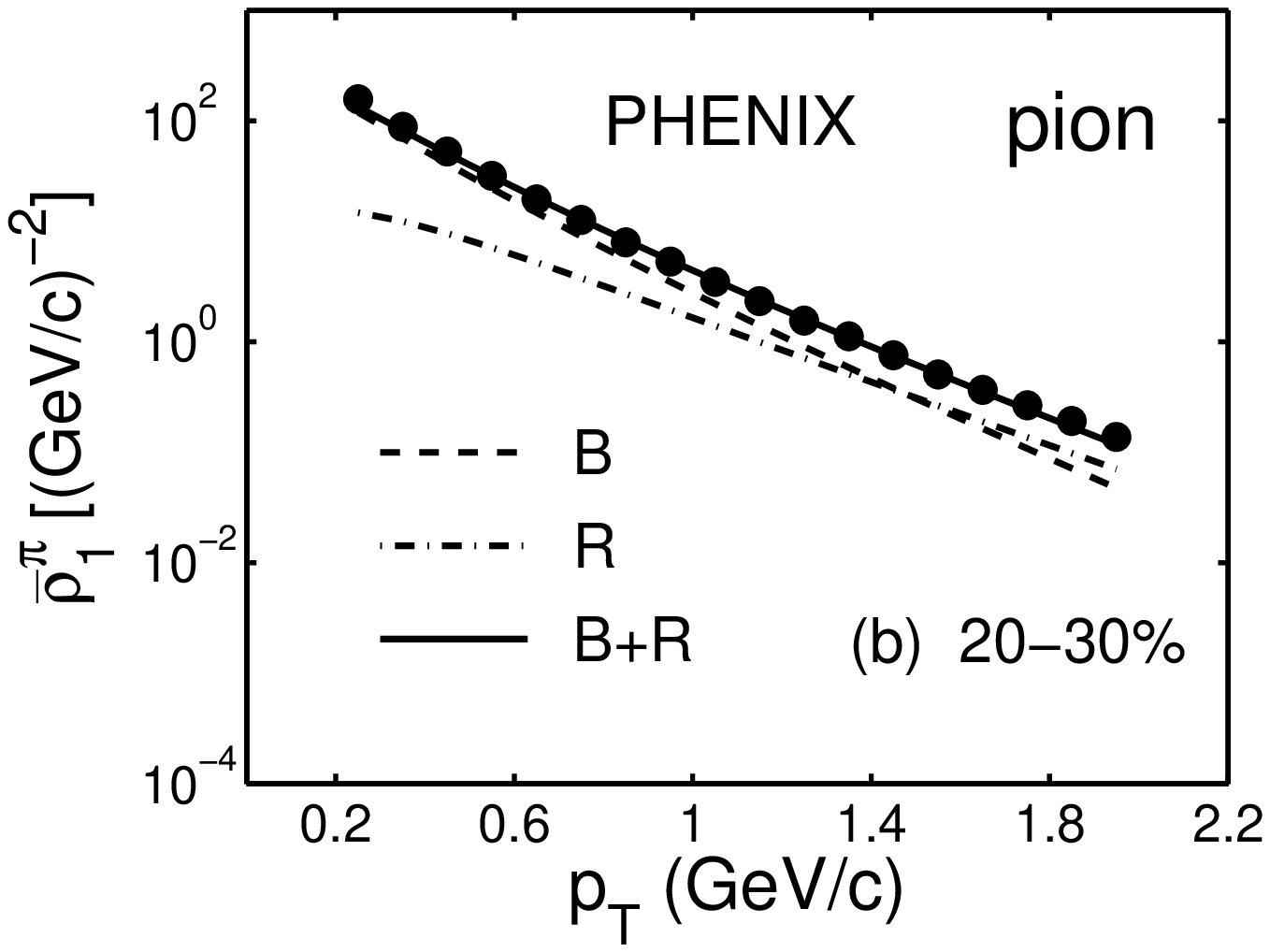}
\caption{Inclusive distributions for pion showing the base (B) component by dashed line and ridge (R) component by dash-dotted line for (a) 0-5\% and  (b) 20-30\%. The solid line is their sum. The data are from Ref.\ \cite{sa}.}
\end{figure}

\noindent Since the exponential factors are the same for $h=\pi$ and $p$, let us consider only the pion \dis\ specifically. In Fig.\ 7 we show $B$ and $R$ components by dashed and dash-dotted lines, respectively, for (a) 0-5\% and (b) 20-30\%. It is in those figures that we exhibit the basic difference between our description of inclusive spectra and those of others. Inclusive ridge represented by $R$ is always present in the single-particle \dis\ whether or not an experiment chooses to do correlation measurement to examine the ridge. Semihard scattering is unavoidable in any nuclear collisions at high energy. Its effect on soft partons is therefore also unavoidable. We quantify the effect by the $R$ component which is determined by the azimuthal anisotropy that is well reproduced in Fig.\ 5. Here in Fig.\ 7 we see it rising above the $\phi$-independent base $B$ component when $\pt$ is higher than 1.4 GeV/c. It is a consequence of the recombination of enhanced thermal partons. For $\pt>3$ GeV/c in addition to the inclusive ridge the  jet component of the semihard partons themselves manifests in the spectra in the form of thermal-shower recombination that characterizes the intermediate-$\pt$ region. Thus we have a smooth transition from low- to intermediate-$\pt$ regions by recognizing the importance of the inclusive ridge component.

It is observed that the dash-dotted lines in Fig.\ 7 are not exactly straight because the ridge component is not exponential in $\pt$. However, for $\pt>1$ GeV/c, $\bar R^\pi(\pt,b)$ can be well approximated by pure exponential. 
In Fig.\ 8 we show by the solid line the $\pt$ dependence of $\bar R_0(\pt)$, defined in Eq.\ (\ref{19a}); it is 
 the part of the ridge \dis s $\bar R^h(\pt,b)$ in Eqs.\ (\ref{18}) and (\ref{19}) that is common for $h=\pi$ and $p$ and is independent of $b$. From the values of $T$ and $T_B$ that we now know, we have $\tilde T=2.39$ GeV. The (red) dashed line is a straight-line approximation of the solid curve for $\pt>1$ GeV/c by
\bq
\bar R_0(\pt)\approx R_0 e^{-\pt/T'},   \qquad  T'=0.326\ {\rm GeV} .  \label{36}
\eq
Thus the ridge \dis\ is harder than the inclusive \dis\ characterized by $T=0.283$ GeV. This is a property that is known from triggered ridges \cite{bia}, but now it is for untriggered inclusive ridge.

\begin{figure}[tbph]
\includegraphics[width=.45\textwidth]{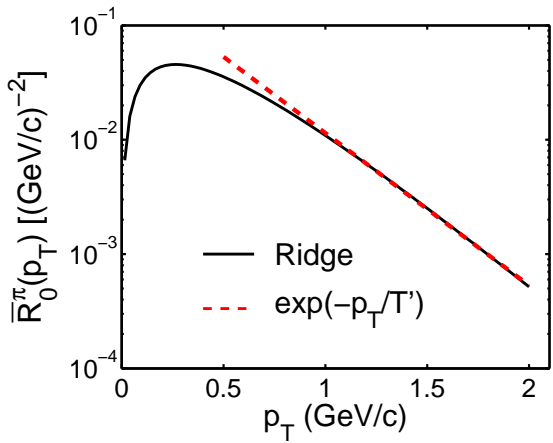}
\caption{(Color online) The $p_T$ dependence of $\bar R_0(\pt)$ defined in Eq.\ (\ref{19a}), represented by the solid line. The (red) dashed line  is a straight-line approximation for $p_T>1$ GeV/c, expressed by Eq.\ (\ref{36}).}
\end{figure}

The enhancement of $T'$ over $T$ is an important point to note. Physically, it means that the ridge is a consequence of the passage of semihard partons through the medium, whose energy losses enhance the thermal partons in the vicinities of the trajectories. We know  that the enhancement factor is $Z(\pt)$, which has the necessary $\pt$ dependence to render $v_2^h(\pt,b)$ to be in good agreement with the quadrupole data. In particular, the property that $Z(\pt)\to 0$ as $\pt\to 0$ is essential to guarantee that $v_2^h(\pt,b)\to 0$ in the  same limit. The effect of $Z(\pt)$ at larger $\pt$ is to increase $T_B$ to $T'$. Although $Z(\pt)$ increases exponentially, its net effect on $\bar R_0(\pt)$ is suppressed by $e^{-\pt/T_B}$. The effective inverse slope $T'$  for $\pt>1$ GeV/c is larger than $T$ of the inclusive by 43 MeV, roughly the same as what Putschke reported on the first discovery of ridge, where the triggered ridge has $T'$ larger than that of the inclusive by 45 MeV \cite{jp}.

There are, however, subtle differences between triggered and untriggered ridges. Experimentally, it is necessary to do correlation measurements to learn about the properties of the ridge, which is extracted by a subtraction scheme. The inverse slope $T'$ can be compared  to the inclusive $T$ of the background. 
For single-particle \dis\ the only measurable inverse slope is $T$ for the inclusive spectrum. 
Theoretically, we assert that ridges do not disappear just because triggers are not used. 
The inverse slope $T'$ for $\bar R_0(\pt)$ cannot be measured directly. It is larger than both $T$ and $T_B$ because $\bar R_0(\pt)$ is the difference between the two exponentials for $\bar\rho_1$ and $B$,
represented by the middle term in Eq.\ (\ref{19a}),  which vanishes as $\pt\to 0$.
A physically more sensible way to compare the various inverse slopes is to recognize that $T'$ is significantly larger than $T_B$ because of the enhancement effect due to semihard scattering, and that $T$ is the effective slope of the inclusive \dis, $B+\bar R$, that is measurable and is between $T_B$ and $T'$.

Although our concern in this paper has been restricted to the midrapidity region, the physics of inclusive ridge can be extended to non-vanishing pseudo-rapidity $\eta$. In Ref.\ \cite{ch} a phenomenological relationship is found between the triggered ridge \dis\ in $\Delta\eta$ and the inclusive \dis\ in $\eta$ with the implication that there is no long-range longitudinal correlation. However, there can be transverse correlation due to transverse broadening of forward (or backward) soft partons as they move through the conical vicinity of the semihard partons. The enhancement of the thermal partons due to energy loss is just as we have described in this paper. Indeed, the term representing the enhanced $\pt$ \dis\ in Ref.\ \cite{ch} is essentially identical to that expressed in Eq.\ (\ref{19a}). Similar consideration has also been used in the explanation of the ridge structure found at LHC \cite{cms,hy2}. In this paper we have presented the most detailed quantitative analysis of the RHIC data in the formalism of inclusive ridge that sets the foundation for the ridges at $|\Delta\eta|>0$.

\section{Conclusion}

Our study of inclusive ridge \dis s has consolidated earlier exploratory work with firm phenomenological support, and therefore succeeded in extending the hadronization formalism from intermediate-$\pt$ region to below 2 GeV/c, exposing thereby an aspect of physics that has not been included in other approaches. The effect of semihard scattering on soft partons is accounted for by the ridge component whose azimuthal behavior is totally characterized by $S(\phi,b)$; it is a calculable quantity bounding the surface segment through which semihard partons can contribute to the formation of a ridge particle at $\phi$. In an earlier paper \cite{hz} we showed the connection between $S(\phi,b)$ and the dependence of the triggered ridge yield on the $\phi_s$ of the trigger angle relative to the reaction plane. Now, we have exhibited the central role that $S(\phi,b)$ plays in determining the azimuthal quadrupole $v_2^h(\pt,b)$ of inclusive \dis s. Thus the inclusive ridge \dis\ that we have advanced in this series of work serves as a bridge between the single-particle \dis\ and the two-particle correlation. Since semihard partons are copiously produced before thermalization is complete, it is an aspect of physics that should not be ignored. The success in fitting $v_2^h(\pt,b)$ for all central and mid-central collisions and for both $\pi$ and $p$ by one  parameter $T_B$ therefore leads to claim of relevance as much as viscous hydro does. 

Another attribute of our approach is to unify the production of pions and protons in one hadronization scheme based on the recombination of enhanced thermal partons so that their spectra have the same inverse slope $T$ despite apparent differences in the low-$\pt$ data. That same scheme when extended to $\pt\sim 3$ GeV/c explains readily the observed large $p/\pi$ ratio. Since the ridge components in our formalism are the same for $\pi$ and $p$, we can predict that the $p/\pi$ ratio is also large in the triggered ridge as in the inclusive.

The property about ridge that most investigators are concerned about is the large $\Delta\eta$ range found in correlation experiments. That is an aspect of the problem that has been addressed in Ref.\ \cite{ch}. Our focus in this paper is on the hidden aspect of the ridge that is not easily detected, but is pervasive because it is in the inclusive \dis. Phenomenological success found in this paper puts the idea on solid footing. If the concept of inclusive ridge is important at RHIC, then its relevance at LHC will be predominant. Since the structure of $v_2^h(\pt,b)$ expressed in terms of $S(\phi,b)$ in Eq.\ (\ref{22}) is independent of initial density, viscosity, or even collision energy, except $\tilde T$, we would expect $v_2$ measured at LHC to be essentially similar  to what is shown in Fig.\ 5 for both $\pi$ and $p$. A preliminary look at the data from ALICE \cite{aam} leads us to believe that such an expectation may not be unrealistic.

\section*{Acknowledgment}
This work was supported,  in part,  by the U.\ S.\ Department of Energy under Grant No. DE-FG02-96ER40972 and by the Scientific Research Foundation for Young Teachers, Sichuan University under No. 2010SCU11090 and Key Laboratory of Quark and Lepton Physics under Grant No. QLPL2009P01.

\end{document}